\documentclass[aps,preprint,amsmath,amssymb]{revtex4-1}
\usepackage{times}
\usepackage{graphicx}
\usepackage{subfigure}
\usepackage{color}
\usepackage{bm}

\begin{document}

%\newcommand{\jp}[1]{\textcolor{red}{#1}}

%\title{Kondo Destruction and Quantum Criticality in Kondo Lattice Systems}

\begin{center}

{\large\bf Kondo Destruction and Quantum Criticality in Kondo Lattice Systems}
\\[0.5cm]

%\author{Qimiao Si$^{1,}$\thanks{E-mail address: qmsi@rice.edu},
%J. H. Pixley$^{1}$, 
%Emilian Nica$^{1}$, Seiji J. Yamamoto$^{1}$, \\
%Pallab Goswami$^{2}$, Rong Yu$^{3}$,  Stefan Kirchner$^{4,5}$
%}

%% Notice placement of commas and superscripts and use of &
%% in the author list

Qimiao Si$^{1}$, 
J. H. Pixley$^{1}$, 
Emilian Marius Nica$^{1}$, 
Seiji J. Yamamoto$^{1}$, \\
Pallab Goswami$^{2}$, 
Rong Yu$^{3}$, 
Stefan Kirchner$^{4,5}$, 

{\em $^1$Department of Physics and Astronomy, Rice University,
Houston, TX 77005, USA}
\\

{\em $^2$National High Magnetic Field Laboratory, Florida State University, \\
Tallahassee, FL 32310, USA}\\

{\em $^3$Department of Physics, Renmin University of China, Beijing, 100872, China}\\

{\em $^4$Max Planck Institute for the Physics of Complex Systems, 01187 Dresden, Germany}\\

{\em $^5$Max Planck Institute for Chemical Physics of Solids, 01187 Dresden, Germany}\\

%%\maketitle

%%\begin{abstract}
%%\vspace{0.5cm}
\end{center}

\vspace{0.5cm}
{\bf
Considerable efforts have been made in recent years to theoretically understand
 quantum phase transitions in Kondo lattice systems. 
A particular focus is on Kondo destruction, which leads to quantum criticality that goes beyond the 
Landau framework of order-parameter fluctuations. This unconventional quantum criticality has provided an
understanding of the unusual dynamical scaling observed experimentally. It has also predicted 
 a sudden jump of the Fermi surface and an extra (Kondo destruction) energy scale, both of which 
have been verified by systematic experiments. Considerations of Kondo destruction have in addition
yielded a global phase diagram, which has motivated the current interest in 
heavy fermion materials with variable dimensionality or geometrical frustration. Here we summarize
these developments, and discuss some of the ongoing work and open issues. We also consider the 
implications of these results for superconductivity.
Finaly, we address the effect
of spin-orbit coupling on the global phase diagram,  suggest that SmB$_6$ under pressure may display unconventional
superconductivity in the transition regime between a Kondo insulator phase and an antiferroamgnetic metal phase,
and argue that the interfaces of heavy-fermion heterostructures 
will provide a fertile setting to explore topological properties of both
Kondo insulators and heavy-fermion superconductors.
}
%%\end{abstract}

\newpage

\section{Introduction}

Quantum criticality is currently being studied in a wide variety of strongly correlated
electron systems.
It provides a mechanism for both non-Fermi liquid excitations and unconventional superconductivity. 
Heavy fermion metals represent a prototype system to study the nature 
of quantum criticality, as well as the novel phases that emerge in the vicinity of a quantum 
critical point (QCP) \cite{JLTP-issue10,Si-Science10}.

Over the past decade, Kondo lattice systems have provided a setting for extensive theoretical analysis
of quantum phase transitions between ordered antiferromagnetic (AF) and paramagnetic ground states.
Various studies have revealed a class of unconventional QCPs that goes beyond the Landau framework
of order-parameter fluctuations. This local quantum criticality incorporates the physics of Kondo destruction. 
Considerations of unconventional quantum criticality have naturally led to the question of the role
of Kondo destruction in the emergent phases. Consequently, a global phase diagram has recently been proposed.

In this article, we give a perspective on this subject and discuss the recent developments. We also point out 
several outstanding issues and some new avenues for future studies.

\section{Quantum criticality}

A quantum many-body Hamiltonian may contain terms that lead to competing ground states. 
A textbook example \cite{Pfeuty.70,Young.75}
is the problem of a chain of Ising spins,  containing both 
a nearest-neighbor ferromagnetic exchange interaction 
between the spins and a magnetic field applied along a transverse direction. 
The exchange interaction 
favors a ground state in which all the spins are aligned, which spontaneously 
breaks a global $Z_2$  symmetry and yields the familiar ferromagnetic order. The transverse field,
on the other hand, prefers a ground state in which all the spins point along transverse direction;
this state does not spontaneously break any symmetry of the Hamiltonian, 
and is therefore a magnetically-disordered state. 

\begin{figure}[t!]
\begin{center}
\includegraphics[width=5.0in]{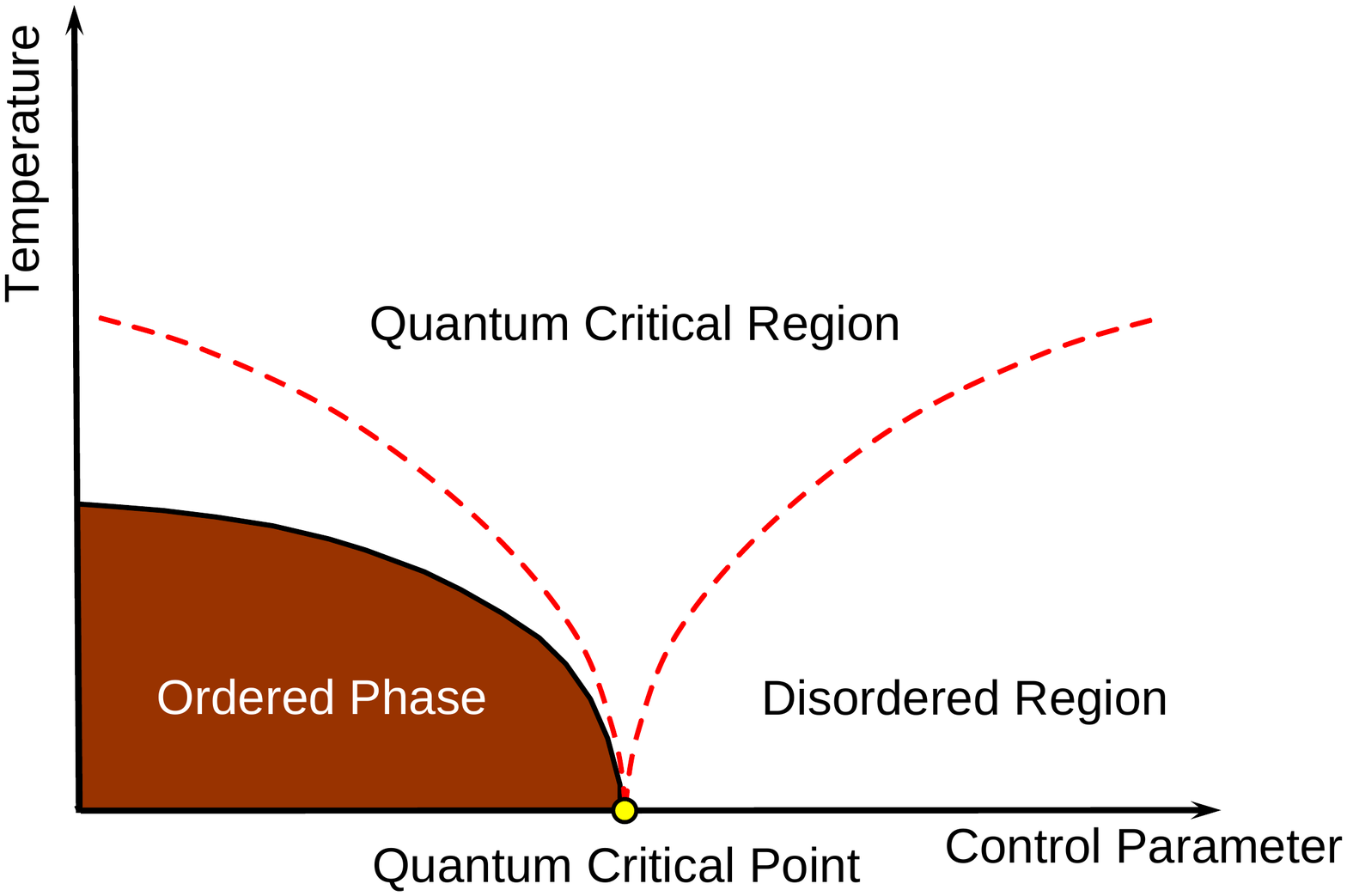}
\end{center}
\caption{(Color online) Quantum critical behavior in the generic phase diagram 
of temperature and a non-thermal control parameter.
}
\label{fig:qcp}
\end{figure}

In general, the ratio of such competing interactions specifies a control parameter, which 
tunes the system from one ground state to another
through a quantum phase transition. A typical case is illustrated in Fig.~\ref{fig:qcp}, 
where the quantum phase transition
goes from an ordered state to a disordered one. When it is continuous, the transition occurs at a QCP.

%%% Figure 1%%%%
%%%%%%%%%%%%

In the Landau framework, the phases are distinguished by an order parameter, which characterizes 
the spontaneous symmetry breaking. The quantum criticality is then described in terms of
$d+z$-dimensional fluctuations of the order parameter in space and time. 
Here, $d$ is the spatial dimension and $z$ is the dynamic exponent.

For weak metallic antiferromagnets, the magnetization associated with the ordering wavevector
characterizes a spin-density-wave (SDW) order. The QCP separates the SDW phase from a paramagnetic 
Fermi liquid state. The collective fluctuations are described in terms of a $\phi^4$ 
theory of order-parameter 
fluctuations~\cite{Hertz.76}.

\begin{figure}[t!]
   \centering
    \includegraphics[width=3.5in]{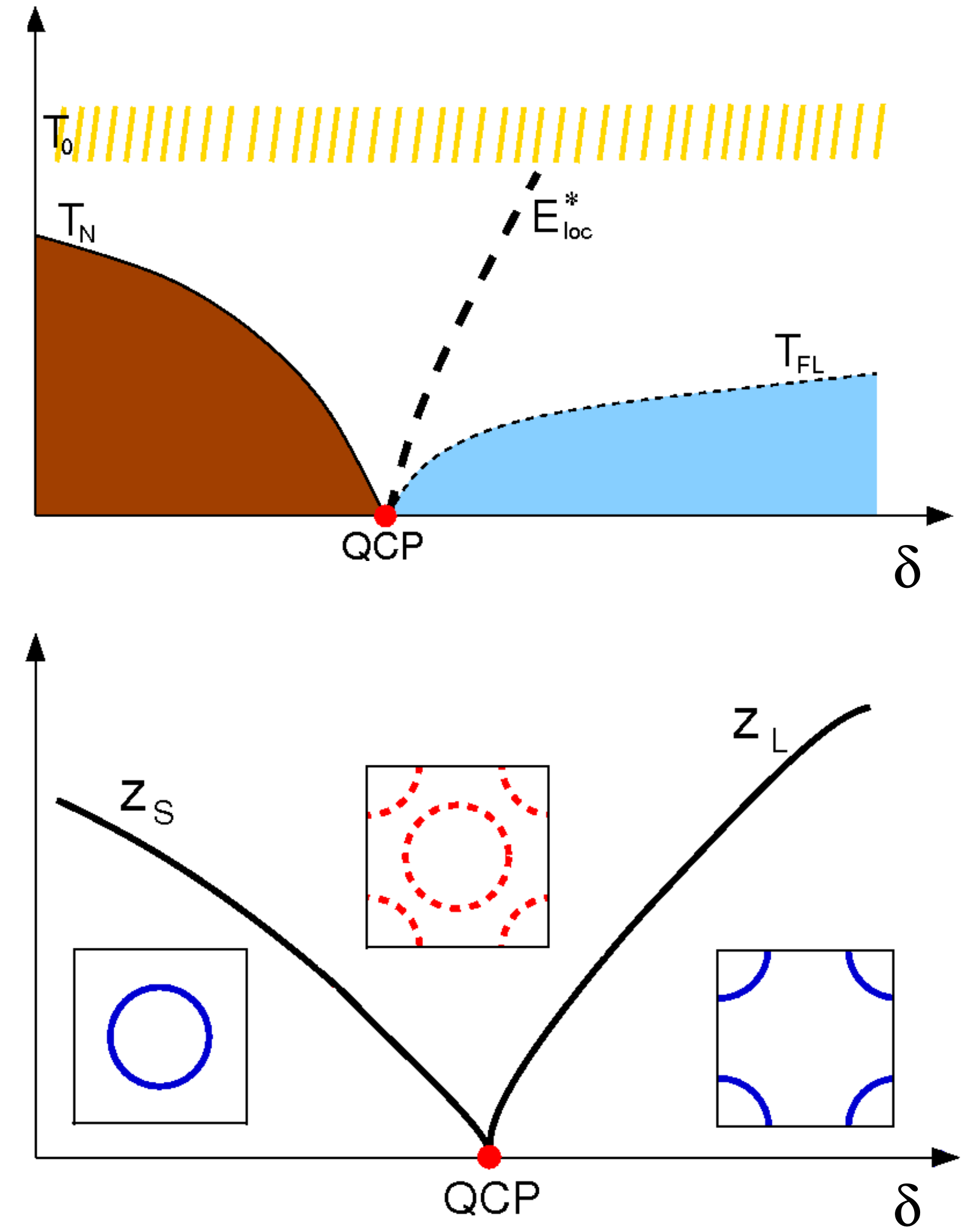}
   \caption{(Color online) Local quantum criticality (top panel) and the corresponding
  $\delta$-dependence of the quasiparticle spectral weights $z_S$ and $z_L$, respectively
   for small and large Fermi surfaces (bottom panel). Here, $\delta \equiv T_K^0/I$ is the control parameter, 
   % is  $\delta \equiv T_K^0/I$, the ratio of the bare Kondo energy
  % scale to the RKKY interaction energy scale. 
 and $T_0$ 
  marks the initial onset of Kondo screening process;
   $T_N$ and 
   $T_{FL}$ are respectively the N\'eel and Fermi-liquid temperatures. 
   $E_{\mathrm{loc}}^*$ characterizes the Kondo destruction, 
   separating the part of the phase diagram where the system flows
   towards a Kondo-singlet ground state from that where the flow is towards a Kondo-destroyed ground state.
   The bottom panel also illustrates the small (left) and large (right) Fermi surfaces, and the fluctuating Fermi surfaces
   (middle) associated with the QCP.
   }
   \label{fig:lqcp}
\end{figure}

In heavy fermion metals, QCPs between an AF phase and a paramagnetic heavy-fermion state
have been observed 
in a number of compounds \cite{JLTP-issue10,Si-Science10}.
The local quantum criticality (Fig.~\ref{fig:lqcp})
has new critical modes associated with the destruction of the Kondo effect,  
in addition to the fluctuations of the AF order parameter \cite{Si-Nature,Colemanetal}. 
It has provided an understanding of unusual dynamical scaling properties observed in quantum critical heavy
fermion metals  \cite{Schroder,Aro95.1}, and made predictions regarding the evolution of Fermi surfaces
and emergence of new energy
scales that have been verified by subsequent experiments in YbRh$_2$Si$_2$ and 
CeRhIn$_5$ \cite{Pas04.1,Shi05.1,Geg07.1,Fri10.2}.

%%% Figure 2%%%%
%\begin{figure}[htbp]
%%%%%%%%%%%%

\section{From the Kondo Effect to its Destruction}

\subsection{Kondo effect}

The Kondo effect was originally studied in the context of a single-impurity Kondo model:
\begin{eqnarray}
H_{\rm Kondo}=
 H_0
+  J_K {\bf S} \cdot {\bf s}^c_{0}
\quad .
\label{kondo-lattice-model}
\end{eqnarray}
Here, $H_0= \sum_{ {\bf k},\sigma} \varepsilon _{\bf k} 
c^{\dagger}_{ {\bf k} \sigma}
c^{\phantom\dagger}_{{\bf k} \sigma} $, ${\bf s}^c_{0} =  c^{\dagger}_{0\sigma}
(\bm{\tau}_{\sigma\sigma'}/2) c^{\phantom\dagger}_{0\sigma'}$, with
$\bm{\tau}$ denoting a vector of Pauli matrices,  and $c^{\dagger}_{0\sigma}$ creates an electron of spin $\sigma$ 
at the impurity site $0$;
the Kondo couling $J_K$ is antiferromagnetic ($J_K>0$). 
The renormalization-group (RG) beta-function is, to quadratic order \cite{Hew97.1}:
\begin{eqnarray}
\frac{d J_K}{d l} \equiv \beta(J_K) = J_K^2 \quad .
\label{beta-function-Kondo}
\end{eqnarray}
The positive sign implies that the effective Kondo coupling grows as the energy is lowered.
The RG flow is towards a strong-coupling fixed point, which controls the physics below a bare Kondo energy 
scale: 
%\begin{eqnarray}
%T_K^0 \approx \rho_0^{-1} \exp \left (
%-1/\rho_0 J_K \right ) \quad ,
%\label{kondo_temperatuer}
%\end{eqnarray}
$T_K^0 \approx \rho_0^{-1} \exp \left (
-1/\rho_0 J_K \right )$,
where $\rho_0$ is the density of states of the conduction electrons at the Fermi energy.
At the strong-coupling fixed point, the local moment and the spins of the conduction electrons
are locked into an entangled singlet state:
\begin{eqnarray}
|{\rm Kondo~singlet}\rangle=\frac{1}{2}(|\!\uparrow\rangle_f|\!\downarrow\rangle_{c,FS}
-|\!\downarrow\rangle_f|\!\uparrow\rangle_{c,FS} ) \quad ,
\label{kondo-singlet}
\end{eqnarray}
where $|\sigma\rangle_{c,FS}$ represents a linear combination of the conduction-electron
states close to the Fermi energy.

This singlet ground state supports a resonance in the low-energy excitation spectrum. The Kondo resonance 
corresponds to a low-energy electronic excitation, which can clearly be seen in
an analysis of the strong-coupling limit, when $J_K$ is taken to be larger than 
the bandwidth of the conduction
electrons, and has been readily described in terms of a slave boson method \cite{Hew97.1}.
The Kondo coupling is converted 
into an effective hybridization, $b^*$, between an emergent fermion $f_{\sigma}$ and the conduction electrons.

\subsection{Kondo lattice and heavy Fermi liquid}

In stoichiometric heavy fermion compounds containing, {\it e.g.}, Ce or Yb elements,
the partially-filled 4$f$ electrons are strongly correlated. They behave as a 
lattice of effective spin-$1/2$ local moments, which describe the 
magnetic degrees of freedom of the lowest Kramers-doublet 
%ground state 
atomic levels. 
This yields a Kondo lattice Hamiltonian:
\begin{eqnarray}
H_{\rm KL}=
H_0
 +
 \sum_{ ij } I_{ij}
{\bf S}_i \cdot {\bf S}_j
+  \sum_{i} J_K {\bf S}_i \cdot {\bf s}^c_i .
\label{kondo-lattice-model}
\end{eqnarray}
The Kondo coupling $J_K$ is antiferromagnetic and we will focus on an antiferromagnetic RKKY interaction,
$I_{ij}>0$.

At high energies, the local moments are essentially decoupled, and Eq.~(\ref{beta-function-Kondo}) 
would continue to apply,
signifying the 
initial development of Kondo 
%scaling as the energy/temperature is lowered. 
screening process.
What happens in the ground state, however, will depend on the competition between the Kondo and RKKY interactions.

%%%% Figure 3%%%%
%\begin{figure}[htbp]
\begin{figure}[t!]
   \centering
    \includegraphics[width=5.0in]{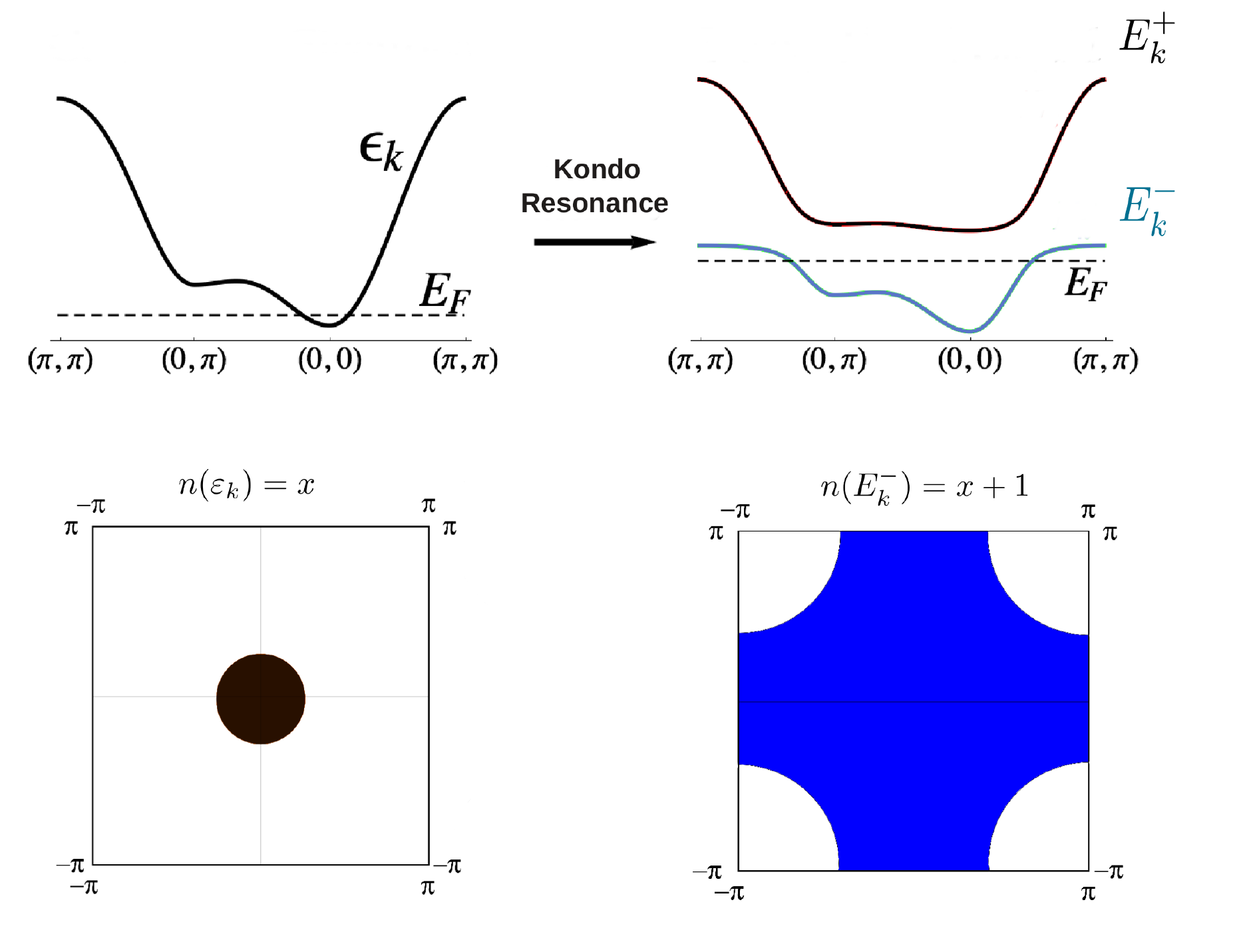}
   \caption{(Color online)  (Left) The energy dispersion of the conduction-electron band. The Fermi surface
   is small in that it only involves the $x$ conduction electrons per unit cell.
  (Right) The bands of the hybridized heavy Fermi liquid. A Kondo/hybridization gap separates the two bands.
  The Fermi surface is large in that it also counts the local moments. Without a loss of generality,
  we have taken $0<x<1$.
   }
   \label{fig:hfl}
\end{figure}
%%%%%%%%%%%%%

Consider first the regime where the Kondo effect dominates, with  $T_K^0$ being much larger 
than the RKKY interaction.
The physics of this regime can be inferred by taking the bare Kondo coupling $J_K$ to be greater 
than the 
bandwidth $W$ of conduction electrons~\cite{Si10.2,Lac85.1,Noz98.1}. 
The Fermi surface will be large, enclosing 
$1+x$ electrons per unit cell. When $J_K/W$ is reduced to being 
considerably smaller than $1$, while keeping $I/T_K^0$
%negligibly 
small, continuity dictates that the entangled Kondo singlet state still characterize 
the ground 
state, and the Fermi surface will remain large. This can be seen, microscopically, through the slave-boson approach~\cite{Auerbach-prl86,Millis87,Burdin.00}.
The Kondo resonance in the excitation spectrum appears as 
a pole in the conduction-electron self-energy:
\begin{eqnarray}
\Sigma({\bf k},\omega)
=\frac{(b^*)^2}{\omega-\varepsilon_f^*} \quad ,
\label{sigma-pole}
\end{eqnarray}
where the self energy is defined through the Dyson
equation:
%\begin{eqnarray}
%G_c({\bf k},\omega) = \frac{1}
%{\omega-\varepsilon_{\bf k} -
%\Sigma({\bf k},\omega)} \quad .
%\label{gc-Dyson-equation}
%\end{eqnarray}
$G_c({\bf k},\omega) =
\left [ \omega-\varepsilon_{\bf k} -
\Sigma({\bf k},\omega) \right ]^{-1}$.
The conduction-electron Green's function now has two poles,
%\begin{eqnarray}
%G_c({\bf k},\omega) =
%\frac{u_{\bf k}^2}{\omega-E^+_{{\bf k}}}
%+
%\frac{v_{\bf k}^2}{\omega-E^-_{{\bf k}}} \quad ,
%\label{gc-two-poles}
%\end{eqnarray}
%with 
at energies
\begin{eqnarray}
E^{\pm}_{{\bf k}} &=& (1/2)[\varepsilon_{\bf k}+\varepsilon_f^*
\pm \sqrt{(\varepsilon_{\bf k}-\varepsilon_f^*)^2+4(b^*)^2} ] 
\quad ,
\label{gc-two-poles2}
\end{eqnarray}
%The coherence factors are specified by $u_k^2+v_k^2=1$ and 
%$u_k^2 - v_k^2 = (\varepsilon_{\bf k} - \varepsilon_f^*)/2\sqrt{(\varepsilon_{\bf k}-\varepsilon_f^*)^2+4(b^*)^2}$.
%$E^{\pm}_{{\bf k}}$ 
which 
%describe 
describe the heavy-fermion bands.
As illustrated in Fig.~ \ref{fig:hfl}, the nonzero $b^*$ describing the Kondo resonances is directly responsible 
for a large Fermi surface.
The quasiparticle residue goes as $z_L \propto (b^*)^2$ (Fig.~\ref{fig:lqcp}).

In addition to the Kondo coupling between the local moments and conduction electrons, the Kondo-lattice Hamiltonian
also contains an RKKY interaction among the local moments. In Eq.~(\ref{kondo-lattice-model}), this has been
explicitly incorporated. The RKKY interaction $I$ represents an energy scale that competes against $T_K^0$, 
the Kondo scale \cite{Doniach,Varma76}.
We define the ratio of the two energy scales, $\delta \equiv T_K^0/I$, to be the tuning parameter.

Historically, the beginning of the heavy-fermion field focused attention on the Fermi liquid behavior
highlighted by a large carrier mass, as well as the exploration of unconventional superconductivity.
It was gradually realized that the Fermi liquid description can break down \cite{Maple.94,Ste01.1}.
In the modern era, there is a wide recognition that quantum criticality underlies the non-Fermi liquid 
behavior in many, if not all, heavy-fermion systems.

\subsection{Quantum Criticality: From Landau approach to Kondo destruction}

To study quantum criticality in  heavy-fermion metals, one would normally assume that 
the Kondo effect 
remains intact across the QCP. The ordered state is then a SDW and the QCP follows the Landau approach
introduced by Hertz for weak antiferromagnets \cite{Hertz.76,Millis,Moriya}.
Because the dynamic exponent in this approach is $z=2$, the effective dimension
 $d+z$ is larger than or equal to $4$ (the upper critical dimension of the $\phi^4$ theory).  
 Therefore, the fixed point is Gaussian.

The search for beyond-Landau quantum criticality 
has focused on the phenomenon of Kondo destruction, 
from which emerges the
inherently quantum modes that do not connect with any spontaneous broken symmetry.

To study  the Kondo destruction, it is important to analyze the dynamical competition between the RKKY and Kondo
interactions.
The RKKY interaction induces dynamical
correlations among the local moments, which are detrimental to the formation of the Kondo singlets.
 The important question is whether this effect is sufficiently
strong to destroy the amplitude of the static singlet, $b^*$, and correspondingly drive the 
energy scale  ($E_{\mathrm{loc}}^*$) 
or temperature scale ($T_{\mathrm{loc}}^*=E_{\mathrm{loc}}^*/k_B$, where $k_B$ is the Boltzmann constant)  to zero.
Such Kondo destruction was discussed
 early on in RG analyses of models containing Kondo-lattice-type effects \cite{Si96.1,SmithSi,Sengupta,Si.99}.
 A systematic study became available in Refs.~ \cite{Si-Nature,Si-prb.03} 
which provided an understanding of the anomalous spin dynamics measured \cite{Schroder}
in the quantum critical heavy fermion metal
CeCu$_{5.9}$Au$_{0.1}$.  
These studies also predicted
the collapse of the 
Kondo-destruction energy scale $E_{\mathrm{loc}}^*$, a sudden jump of the 
Fermi  surface across the QCP, and the critical nature of the quasiparticles  on the whole Fermi surface
at
the QCP (cf.\ Fig.\,\ref{fig:lqcp}). This approach was
connected to some general theoretical considerations \cite{Colemanetal}. 
The Kondo destruction has since also been
studied in a fermionic slave-particle approach  \cite{Sen04.1,PaulPepinNorman.07}
as well as in dynamical mean field
theory \cite{Deleo.08}. The critical quasiparticles  on the entire Fermi surface
have also been considered using a  self-consistent  method
\cite{WolfleAbrahams_prb11}.

\section{Kondo destruction and Quantum Criticality of Antiferromagnetic Heavy Fermion Metals}

An AF QCP is expected when the control parameter $\delta$ becomes sufficiently small, {\it i.e.},
when the RKKY interaction is large enough.
One microscopic approach that has been playing an important role is 
 the extended dynamical  mean-field  theory (EDMFT)~\cite{Si96.1,SmithSi-edmft,Chitra.00}.
 
\subsection{EDMFT approach}

In the EDMFT approach, the fate of the Kondo effect is studied through the
Bose-Fermi Kondo model,
\begin{eqnarray}
{\cal H}_{\text{imp}} &=& H_{\mathrm{Kondo}} +\sum_{p}
w_{p}\,{\bf \Phi}_{p}^{\;\dagger} \cdot {\bf \Phi}_{p} 
\nonumber \\
&& +  g \sum_{p} {\bf S} \cdot \left( {\bf \Phi}_{p} +
{\bf \Phi}_{-p}^{\;\dagger} \right),
\label{eq:h_imp}
\end{eqnarray}
along with the  self-consistency equations 
$\chi_{{loc}} (\omega)
= \sum_{\bf q} \chi ( {\bf q},
\omega )$,
and $G_{{loc}} (\omega) = \sum_{\bf k} G( {\bf k}, \omega )$.
Associated with Eq.~(\ref{eq:h_imp}) are the Dyson equations:
$M(\omega)=\chi_{0}^{-1}(\omega) + 1/\chi_{\rm loc}(\omega)$
and $\Sigma(\omega)=G_0^{-1}(\omega) - 1/G_{\rm loc}(\omega)$, where
$\chi_{0}^{-1} (\omega) = -g^2 \sum_p 2 w_{p} /(\omega^2 -
w_{p}^2)$
and $G_0 (\omega) = \sum_p 1/(\omega - E_p)$.
The self-consistency equations manifest the spatial dimensionality of the magnetic 
fluctuations through the form of the RKKY density of states 
$\rho_{I} (x) \equiv  \sum_{\bf q} \delta ( x  - I_{\bf q} )$.
Two dimensional magnetic fluctuations correspond to
a $\rho_{I}(x)$ which is nonzero at the lower edge,
as typified by the case:
$\rho_{I}(x) = (1/{2 I}) \Theta(I - | x | )$,
where $\Theta$ is the Heaviside step function.
On the other hand, three dimensional magnetic fluctuations are represented by a
$\rho_{I} (x)$ with a square-root form near the lower edge, as given by the example
of  $\rho_{I} (x) = (2/{\pi I^2}) \sqrt{I^2-x^2}\, \Theta(I - | x | ) $.
 
%%% Figure 4%%%%
%\begin{figure}[htbp]
\begin{figure}[t!]
   \centering
   \includegraphics[width=4.5in]{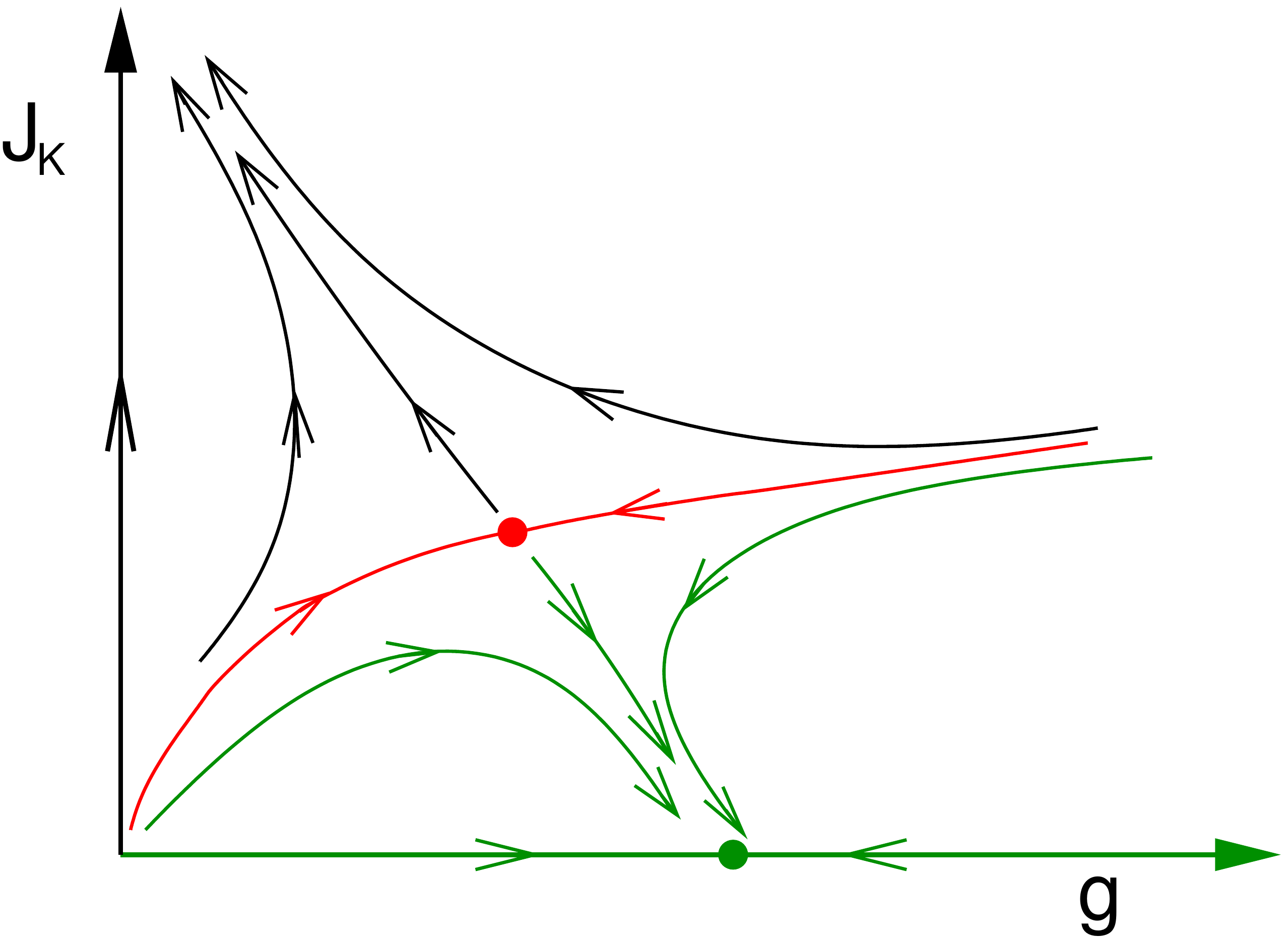}
   \caption{ (Color online) RG flow of the Bose-Fermi Kondo model (with $\epsilon > 0$), showing the 
   strong-coupling Kondo fixed point and its destruction.
   }
   \label{fig:rg_bfkm}
\end{figure}
%%%%%%%%%%%%

\subsection{Kondo destruction}

In the EDMFT approach, the dynamical magnetic correlations of the local moments 
influences the Kondo effect through the bosonic bath. 
Irrespective of the spatial dimensionality, the bosonic
bath has a softened spectrum near the magnetic QCP. Correspondingly, 
it causes an enhanced suppression the Kondo effect.
This effect has been studied extensively,
as in Ref.~\cite{Zhu.04}.  It
 can be most clearly seen through an RG approach of the Bose-Fermi Kondo problem
that utilizes an $\epsilon$-expansion, where $\epsilon$ is defined through the deviation
of the bosonic spectrum from the Ohmic form: $\sum_{p} \delta (\omega -w_p) \sim \omega^{1-\epsilon}$.
The RG equations of the Kondo problem, Eq.~(\ref{beta-function-Kondo}), now have,
to the quadratic order in $J_K$ and $g^2$, 
 the following form  \cite{Si-Nature, ZhuSi, SmithSi,Sengupta,Si.99}:
\begin{eqnarray}
\beta (J_K)  &=& J_K (J_K - g^2 ) 
\nonumber \\
\beta (g) &=& g (\epsilon /2 - g^2) 
\label{eq:rg_bfkm}
\end{eqnarray}
The RG equations yield a Kondo-destruction critical point, as shown in Fig.~\ref{fig:rg_bfkm}.
Importantly,  the zero-temperature local spin susceptibility has the form:
\begin{equation}
\chi_{\mathrm{loc}}(\tau)\sim\tau^{-\epsilon}
\end{equation}
Importantly, the result for the critical exponent is valid to infinite orders in $\epsilon$.

\subsection{Local quantum critical point}

The EDMFT equations have been studied in some detail in a number
of analytical and numerical studies
\cite{Si-Nature,Si-prb.03,GrempelSi,ZhuGrempelSi,SunKotliar.03,Glossop07,Zhu07}.
Irrespective of the spatial dimensionality, the weakening of the
Kondo effect is seen through the reduction of the $E_{\mathrm{loc}}^*$ scale as we approach the 
QCP from the paramagnetic side. 

$E_{\mathrm{loc}}^*$ 
vanishes at the QCP for two-dimensional magnetic fluctuations.
The dynamical spin susceptibility satisfies the following dynamical scaling:
\begin{eqnarray}
\chi({\bf q}, \omega ) =
\frac{1}{f({\bf q}) + A \,(-i\omega)^{\alpha} \mathcal{M}(\omega/T)} \quad .
\label{chi-qw-T}
\end{eqnarray}
The exponent $\alpha$ is found to be near
to 0.75 (between 0.72 and 0.83 derived from different approaches) 
\cite{GrempelSi,Zhu07,Glossop07}.

For three-dimensional magnetic fluctuations, $E_{\mathrm{loc}}^*$
 is reduced but remains non-zero at the QCP. It however terminates inside
the ordered portion of the phase diagram (see below).

In both cases, the zero-temperature transition
is second-order when the effective RKKY interaction appears 
in the same form on both sides of the transition~\cite{SiZhuGrempel.05,SunKotliar.05}.
It is important to stress one effect that is crucial for both the stability of the Kondo-destroyed AF phase 
as well as the second-order nature of the QCP: a dynamical Kondo effect still operates
in the Kondo-destroyed AF phase. We will expound on this point in Sec.~\ref{afs_sdw}.

\section{Global phase diagram}

Kondo destruction and the associated Fermi-surface change represent physics that goes beyond the Landau framework. 
Considerations of
new phases that reflect this physics have led to a 
global phase diagram for the AF Kondo-lattice systems
 \cite{Si.06,Si10.1,Col10.2}.
 This phase diagram was first developed based on theoretical studies
showing the stability of  the ${\rm AF_S}$ phase
\cite{Si.06,Yamamoto.07,Ong_Jones09,Yam10.1}. This is an AF phase with Kondo destruction and an
associated small Fermi surface.

\subsection{Kondo destruction inside antiferromagnetic order: QNL$\sigma$M approach}

Can the Kondo effect be destroyed inside the AF ordered phase?
To address this issue, Ref.~\cite{Yamamoto.07} considered the model with 
SU(2) symmetry, and in the parameter limit 
of $J_K \ll I \ll W$.
(The case with Ising anisotropy is simpler:
because the AF-ordered phase has a spin gap, $J_K$ is irrelevant
and the AF$_S$ phase will be stable.)
We consider the reference limit $J_K=0$ to be 
 the local moments in a collinear AF order, described by a quantum non-linear sigma model 
 (QNL$\sigma$M) and, separately,
the conduction electron band.
The QNL$\sigma$M
\cite{haldane1983,Chakravarty} takes the form:
\begin{eqnarray}
{\cal S}_{\text{QNL}\sigma\text{M}}
=
(c/2g) \int d^dxd\tau\left[ \left(\nabla
{\bf n}
\right)^2 +
c^{-2}\left(
\partial_{\tau}{\bf n}
\right)^2 \right]
\; ,
\label{qnlsm}
\end{eqnarray}
where $c$ is the spin-wave velocity and $g$ measures the amount of quantum fluctuations, which grows as 
the amount of frustration is increased (such as via tuning the ratio of next nearest neighbor to nearest 
neighbor spin-spin interactions on a square lattice).

%%% Figure 5%%%%
%\begin{figure}[htbp]
\begin{figure}[t!]
   \centering
    \includegraphics[width=3.8in]{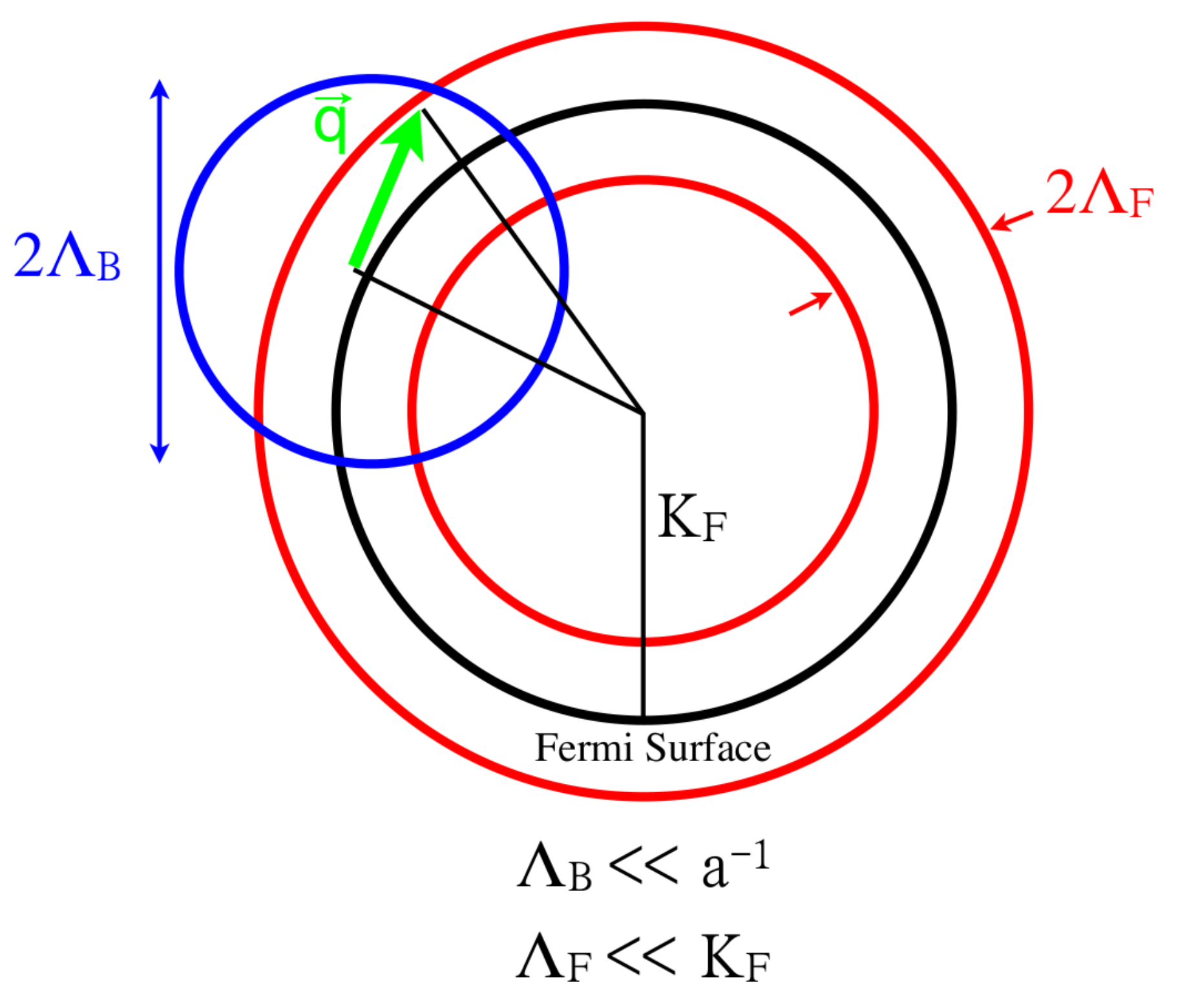}
   \caption{(Color online) Different kinematics in the scaling of the bosonic and fermionic sectors. Figure adapted 
   from Ref.~\cite{Yamamoto09}.
   }
   \label{fig:rg_afs}
\end{figure}
%%%%%%%%%%%%

The case of the AF zone boundary not intersecting the Fermi surface 
of the conduction electrons allows an asymptotically exact analysis.
Expressed in terms of the ${\bf n}$ field of the QNL$\sigma$M,
which represents the staggered magnetization, the
Kondo coupling takes the following form at low energies:
\begin{eqnarray}
{\cal S}_K =
\lambda_K\int d^dxd\tau
~
{\bf s}_c
\cdot
{\bf n} \times \partial_{\tau} {\bf n} .
\label{kondo-lambda}
\end{eqnarray}
While the RG procedure \cite{Yamamoto09} appropriate for combined 
gapless fermionic\cite{shankar1994} and bosonic fields
is in general very involved, the situation simplifies here because the
QNL$\sigma$M has a dynamic exponent $z=1$. 
The kinematics involved in the RG approach is illustrated in 
Fig.~\ref{fig:rg_afs}. The resulting RG equation is:
\begin{eqnarray}
\beta (\lambda_K)  =0
\label{eq:rg_lambdaK}
\end{eqnarray}
In other words, $\lambda_K$ is exactly marginal.
The Kondo coupling 
does not grow, and there is no Kondo singlet formation in the ground state;
{\it i.e.}, the AF phase has a Kondo destruction.
This establishes the stability of the $AF_S$ phase.
A large-$N$ analysis of the low-energy excitations was also carried out in Ref.~\cite{Yamamoto.07}, 
yielding a
self-energy for the conduction electrons:
\begin{eqnarray}
\Sigma({\bf k},\omega)
\propto \omega^d \; .
\label{sigma-no-pole}
\end{eqnarray}
This self-energy lacks a pole, which is to be contrasted with
Eq.~(\ref{sigma-pole}). In other words, the Kondo resonance 
is absent, and the Fermi surface is small.

\subsection{Global Phase Diagram}
\label{sec:global}

\begin{figure}[t!]
   \centering
%\vskip 0.5 cm
\includegraphics[width=4.5in]{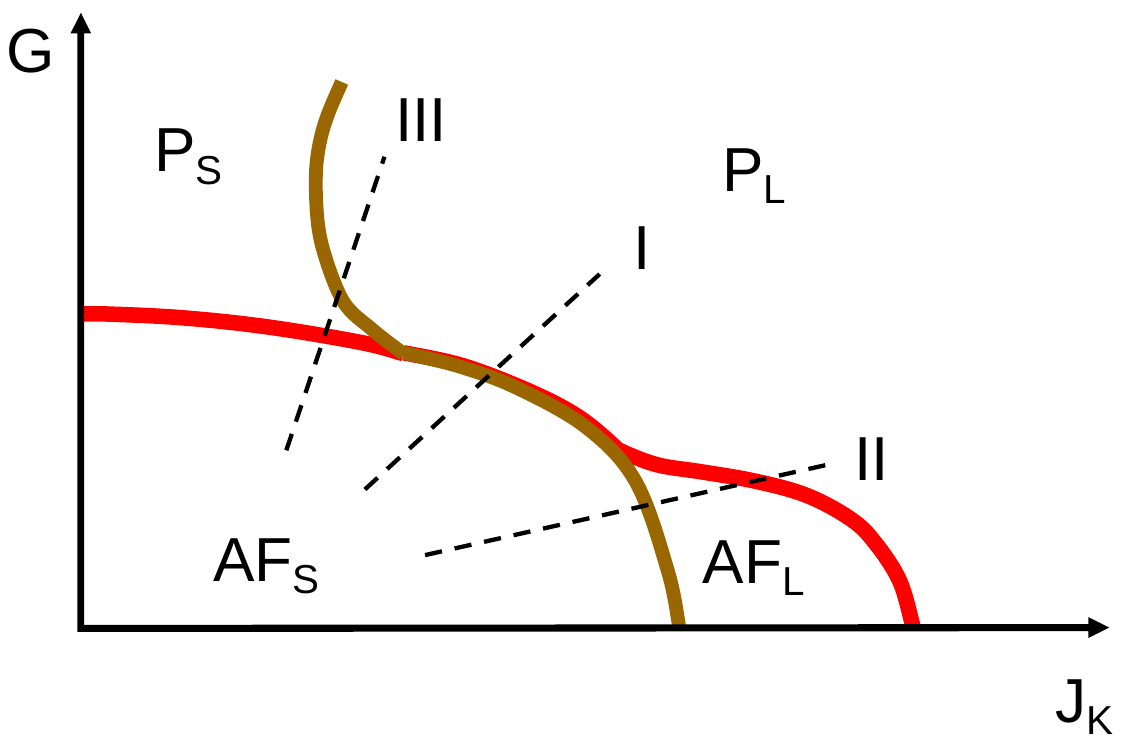}
   \caption{
  The global phase diagram of the AF
Kondo lattice \cite{Si.06,Si10.1}.
 This  $T=0$ phase diagram involves a frustration axis ($G$) and 
 an axis that tunes the Kondo coupling ($J_K$).
 ${\rm P_L}$ and ${\rm P_S}$ are 
paramagnetic phases whose Fermi surfaces are respectively
large and small ({\it i.e.} with or without Kondo resonances).
${\rm AF_L}$ and ${\rm AF_S}$ are their AF counterparts.
   }
   \label{fig:gpd}
\end{figure}

The stability of the $AF_S$ phase raises the question, what are the different possible routes 
to suppress AF order and tune the system from the $AF_S$ phase towards 
the paramagnetic heavy-fermion state ($P_L$ phase)?
The routes are illustrated in the global phase diagram,
Figure~\ref{fig:gpd}. 
This zero-temperature phase diagram involves two parameters:
in addition to 
the  Kondo coupling $J_K$, there is also $G$ which 
measures the degree of the quantum fluctuations of the local-moment magnetism. 
The vertical axis reflects tuning geometrical
frustration or dimensionality.
The Kondo coupling, depicted as the horizontal axis,
is taken to be dimensionless with the conduction-electron bandwidth
$W$ as the normalization factor.
The global phase diagram itself is a two-dimensional projection
of a multi-dimensional phase diagram. In particular,
we have considered the case with a fixed $I/W$ that is considerably smaller 
than $1$. In addition, we have fixed,
$x$, the number of conduction electrons per site, to some non-integer value.

There are three sequences of phase transitions from the 
${\rm AF_S}$ phase to the ${\rm P_L}$ phase.
Trajectory I  represents a direct transition involving Kondo destruction, and
corresponds to the local QCP.
Trajectory II involves an intermediate  ${\rm AF_L}$ phase,
which represents the SDW order of the ${\rm P_L}$ phase.
There is a Kondo-destruction transition inside the AF order,
while the AF to non-magnetic transition is of the SDW type.
Trajectory III involves an intermediate 
 ${\rm P_S}$ phase, which could involve non-magnetic order such 
 as a  valence-bond solid.
 
 %%% Figure 6%%%%
%\begin{figure}[htbp]
%%%%%%%%%%%%

\subsection{Specific cases and multiplicity of tuning parameters}

Our discussion so far is very general. To make further progress, it is important to consider the specific 
cases as well as the specific realizations of the parameter $G$.

One case which is amenable to concrete calculations is the
Ising-anisotropic Kondo problem in the presence of a transverse magnetic field.
As already mentioned in the introduction, the transverse field
introduces quantum fluctuations for the local moments, and provides a means to tune the $G$ axis.
In an EDMFT study, this leads to a Bose-Fermi Kondo model, with Ising anisotropy
and in the presence of a transverse field,
which is supplemented by self-consistency conditions. The transverse-field Bose-Fermi Kondo model
 {\it per se} has 
recently been studied in detail~\cite{Nica}. The calculations have been carried out using 
a version of the numerical renormalization group method~\cite{Glossop.05}, and a line of 
Kondo-destruction fixed points was identified.

Another setting for concrete calculations is the spin-symmetric Kondo lattice model
on the Shastry-Sutherland 
lattice. The parameter $J_2/J_1$ -- the ratio of the exchange interaction on a diagonal bond to that on 
the nearest-neighbor bond -- measures the degree of frustration and is defined as $G$.
A key advantage is that, at large $G$ and $J_K=0$ 
the ground state of the local-moment only model is known exactly
to be a valence-bond solid~\cite{Shastry.81}. A large-N-based calculation~\cite{Pix13.1}
yields a phase diagram that is reminiscent 
of Fig.~\ref{fig:gpd} when the conduction electrons are away from half-filling.

\subsection{Berry phase and the topological defects of N\'eel order}

Considerations of the global phase diagram also opens up the study of the heavy-fermion state
based on the Berry phase and topological defects of local-moment magnetism. This was recently studied 
in the 
QNL$\sigma$M representation of the spin one-half Kondo lattice model on a honeycomb lattice at half filling~\cite{Goswami.13}, (see also Ref.\
~\cite{Goswami.11} for the 1D case).
It has been shown that the skyrmion defects of the antiferromagnetic order parameter host a number of competing 
states.  In addition to the spin Peierls, charge and current density wave order parameters, Kondo singlets also appear
as the competing variables dual to the AF order.  In this basis, the conduction electrons acquire a Berry phase
through their coupling to the hedgehog configurations of the N\'eel order, which cancels the Berry phase
of the local moments. These results demonstrate the competition between the Kondo-singlet formation
and spin-Peierls order when the AF order is suppressed, in a way that is compatible with
 the global phase diagram discussed earlier.

%%% Figure 7%%%%
%%%%%%%%%%%%

\section{Antiferromagnetic order: Kondo destruction vs. spin-density-wave order}
\label{afs_sdw}

We have so far emphasized that the stability of the $AF_S$ phase has been derived based on an (asymptotically exact)
RG analysis. The exact marginality of the Kondo coupling
inside the AF order is to be contrasted with its marginal relevance in the paramagnetic case.
Because the effective Kondo coupling does not grow, the system no longer flows to the 
strong-coupling Kondo fixed point; in the ground state, the static Kondo singlet has zero amplitude.
However, the marginal nature of the effective Kondo coupling
also implies that the Kondo coupling influences the properties at non-zero energies. In other words, a dynamical
Kondo effect operates.

%%% Figure 7%%%%
%\begin{figure}[htbp]
\begin{figure}[t!]
   \centering
   \includegraphics[width=3.8in]{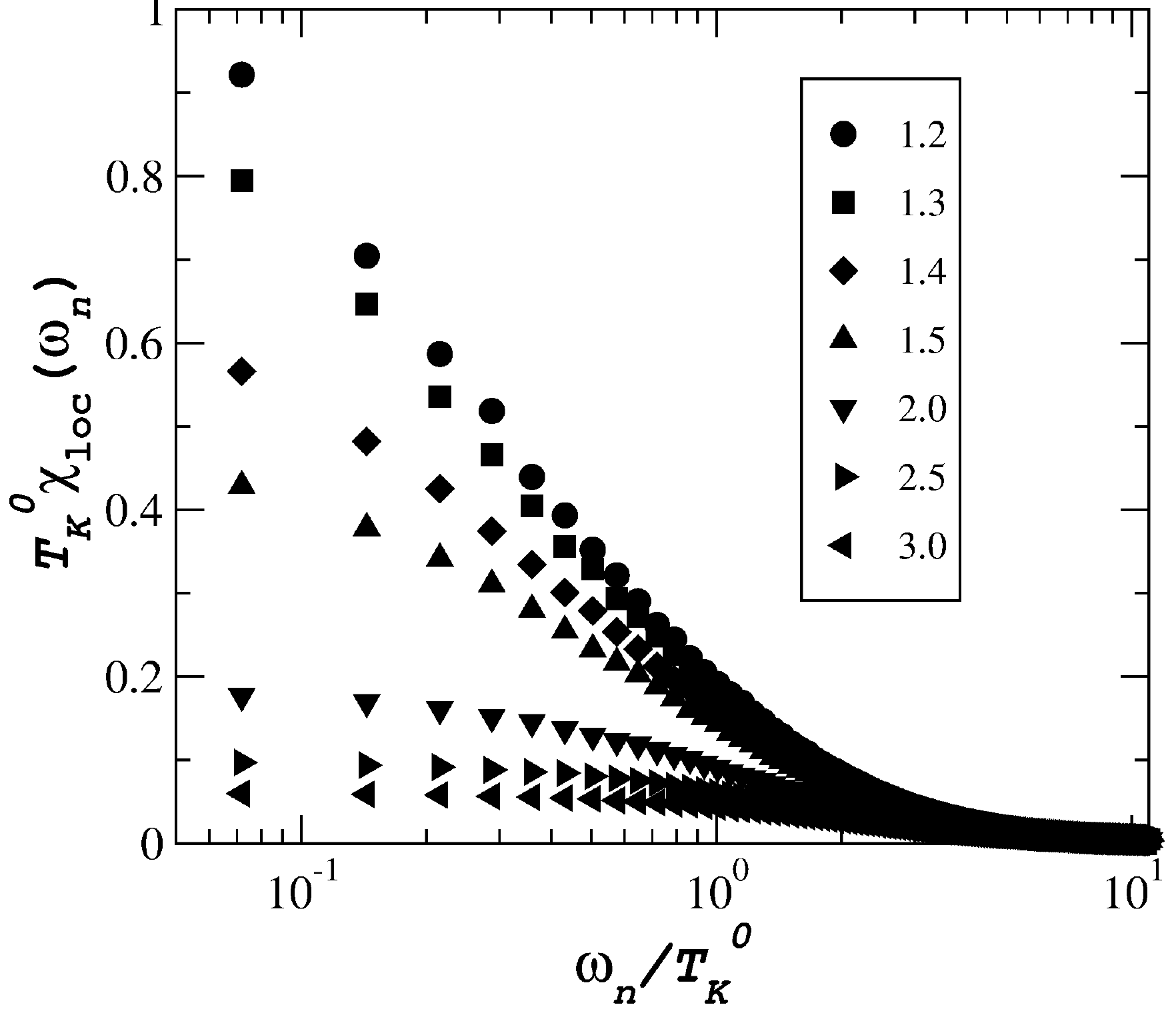}
   \caption{Dynamical Kondo effect in the $AF_S$ phase, for various values of $I/T_K^0$ on 
   the ordered side \cite{ZhuGrempelSi.03}.
   }
   \label{fig:dyn_kondo}
\end{figure}
%%%%%%%%%%%%

This analysis complements the results from the EDMFT studies in the ordered state. Figure~\ref{fig:dyn_kondo} 
shows the local dynamical
spin susceptibility as a function of frequency in the $AF_S$ state~\cite{ZhuGrempelSi.03}. Its increase as $\delta$ 
is tuned towards the QCP reflects the growth of the dynamical Kondo effect.  When $\delta$ reaches $\delta_c$, 
the local dynamics in the ordered state match 
the quantum critical behavior determined from the EDMFT studies in the absence of the order. This demonstrates
the importance of the dynamical Kondo effect in ensuring the second-order nature of the zero-temperature transition.
The same conclusions also emerge in several other studies~\cite{Glossop07,Zhu07}.

%%% Figure 8%%%%
%\begin{figure}[h]
\begin{figure}[t!]
   \centering
   \includegraphics[width=4.8in]{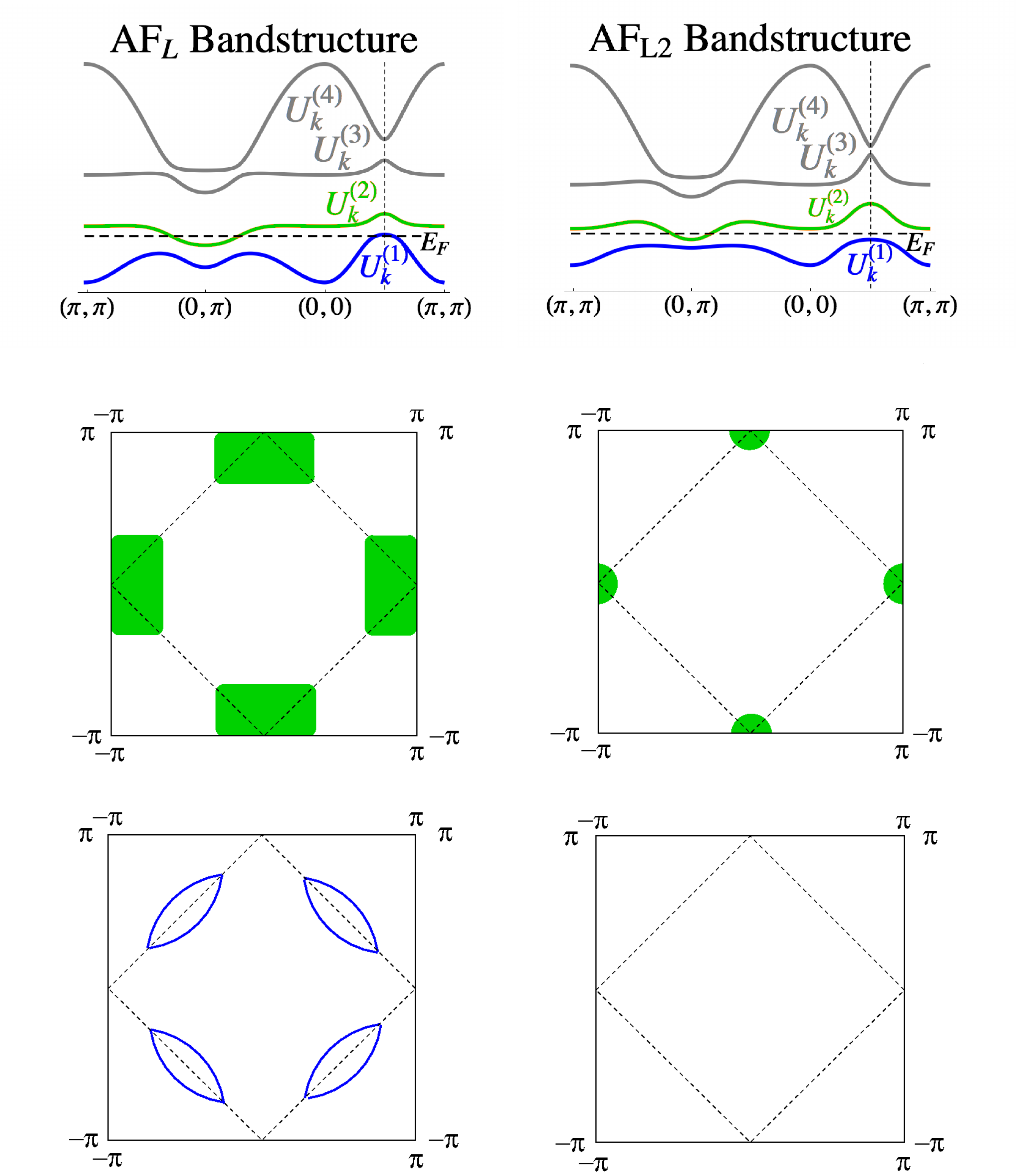}
   \caption{(Color online) $AF_L$ and $AF_{L2}$ phases in the SDW portion of the phase diagram.
   $U_{\bf k}^{\alpha}$, with $\alpha=1,...,4$, are the heavy-fermion bands in the presence 
   of a staggered field associated with a staggered magnetization.
   }
   \label{fig:AFL}
\end{figure}
%%%%%%%%%%%%

Indeed, the dynamical Kondo effect is important for the stability of the $AF_S$ phase. It allows the gain of the Kondo
exchange energy even in the absence of the static Kondo singlet formation. This point is important to understand the 
results of several variational Quantum Monte Carlo studies \cite{Wat07.2,Martin08,Lanata08}.
These studies used a variational wavefunction
for the $AF_S$ phase that not only sets the static Kondo amplitude to zero, but also disallows any Kondo fluctuations 
at finite energies. As such, it cannot energetically compete against the $AF_L$ phase, defined in terms of a variational
wavefunction with a static Kondo-singlet amplitude.
In this way, the approach does not adequately capture the dynamical competition
between the RKKY and Kondo interactions. 
Correspondingly, it is difficult to stablize the $AF_S$ phase. Instead, these studies would only allow the multiple AF ground states with a Lifshtiz transition inside the SDW $AF_L$ phase. In the example shown in Fig.~\ref{fig:AFL},
this corresponds to going from the usual $AF_L$ phase for small AF order parameter, with co-existing electron and hole pockets, to the $AF_{L2}$ phase for larger AF order parameter, in which the hole pocket has disappeared. 
Interestingly, the standard DMFT approach likewise over-emphasizes the Kondo coupling, because
the RKKY interactions do not appear in the dynamical equations 
[the bosonic bath in Eq. (\ref{eq:h_imp}) is absent in DMFT]. The approach therefore 
reduces the regime of stability of the $AF_S$ phase~\cite{Kuramoto.13}. In the terminology of Fig.~\ref{fig:gpd},
it captures the type II transition but misses the type I transition (or, for that matter, the type III transition as well).

\section{Experiments on quantum critical heavy fermions}

Quantum phase transitions in general, and Kondo destruction in particular, have been playing a central role
in the modern studies of heavy fermion magnetism and superconductivity. Here we briefly consider the 
salient properties of the theory that have either been compared to known experiments, or represent
predictions that have been tested by subsequent experiments. More extensive discussions may be found 
in Refs.~\cite{Si.13.1,Steglich.13}.

We have already mentioned the strong evidence \cite{Schroder,Pas04.1,Shi05.1,Geg07.1,Fri10.2}
for local quantum criticality from the heavy-fermion compounds 
CeCu$_{6-x}$Au$_x$, YbRh$_2$Si$_2$, and CeRhIn$_5$. This concerns 
the anomalous
dynamical scaling, an extra energy scale and a sudden
jump of the Fermi surface. Additional evidence has come
 from transport measurements \cite{park-nature06,Par08.1,Knebel.08} in CeRhIn$_5$
 and NMR studies of YbRh$_2$Si$_2$ \cite{Ish02.1}.

The proposed global phase diagram has helped understand
 a surprisingly rich zero-temperature phase
diagram of the Ir- and Co-substituted YbRh$_2$Si$_2$ \cite{Fri09.1,Cus10.1}.
Likewise, it may also provide a means to understand the variety of quantum phase transitions 
under the multiple tuning parameters in CeCu$_{6-x}$Au$_x$ \cite{Sto07.1} and CeRhIn$_5$~\cite{Jiao.13}.

The global phase diagram has suggested that increasing dimensionality
tunes the occurrence of Kondo destruction from
at the onset of AF order to inside the ordered region, which is consistent with the measurements in 
Ce$_3$Pd$_{20}$Si$_6$ \cite{Cus12.1}.

Finally, it also suggests that heavy-fermion materials with lattices that host 
geometrically-frustrated magnetism would be particularly instructive in exploring the upper portion
of the phase diagram, {\it i.e.} the region where the local-moment component contains
especially strong quantum fluctuations. This has provided the motivation for recent studies of heavy-fermion
 metals on Shastry-Sutherland \cite{Kim-prl.13}, Kagome \cite{Fritsch12}, 
 fcc \cite{Mun.13} and triangular~\cite{Khalyavin.13}
 lattices.

%%% Figure 9%%%%
%\begin{figure}[htbp]
\begin{figure}[t!]
   \centering
   \includegraphics[width=5.5in]{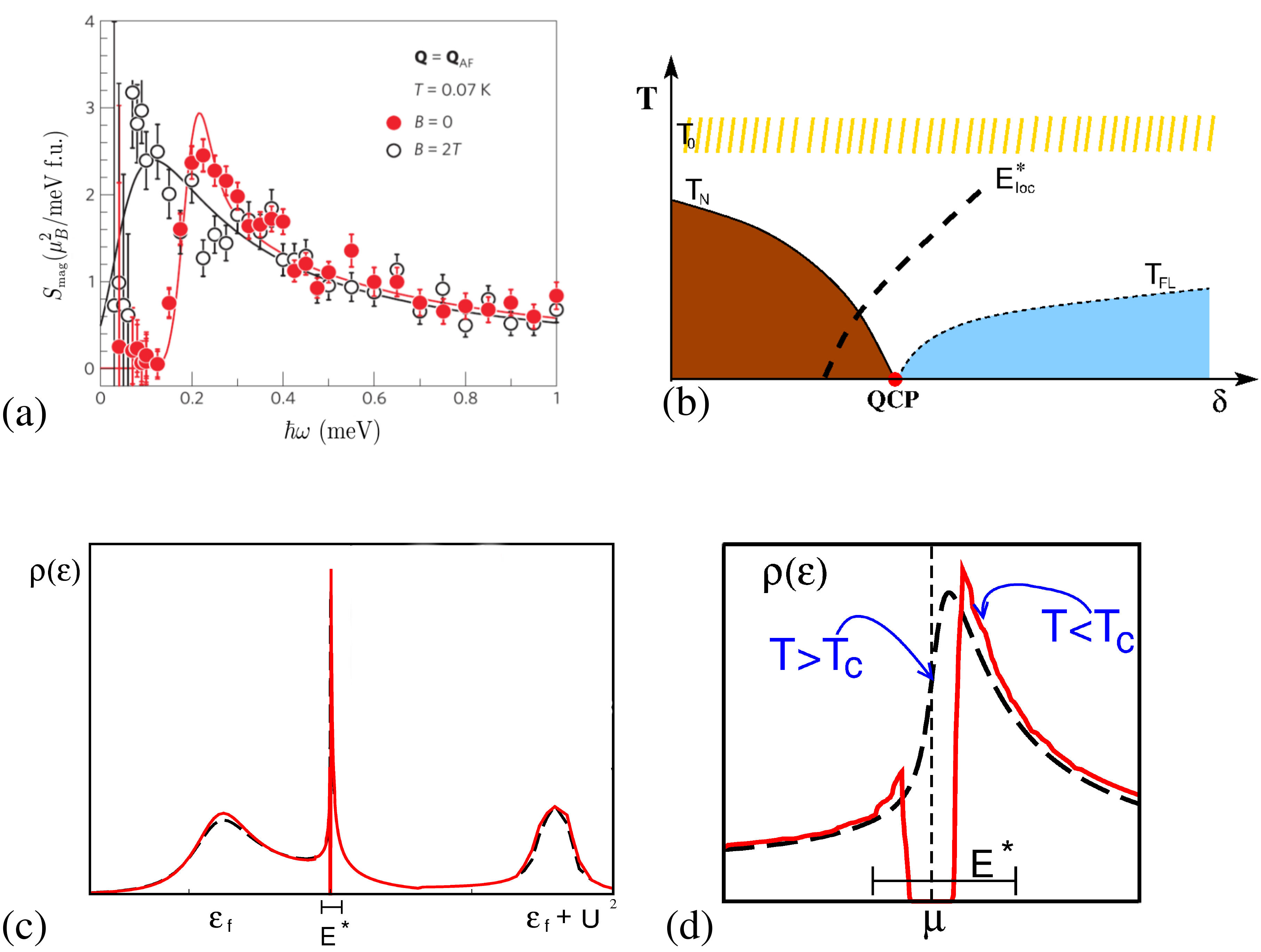}
   \caption{(Color online) (a) Dynamical spin structure factor in the normal and superconducting states 
   of CeCu$_2$Si$_2$, from Ref.~\cite{Stockert-natphys11} (b)(c)(d) Interpretation of the large exchange 
   energy gain in tersm of a nonzero but small Kondo-destruction energy scale $E^*_{\mbox{\tiny loc}}$,
   as introduced in Refs.~\cite{Stockert-natphys11,Stockert-jpsj12}, and described in the main text.
   }
   \label{fig:SC-CeCu2Si2}
\end{figure}
%%%%%%%%%%%%

\section{Implications for superconductivity}

Unconventional superconductivity often arises in the vicinity of magnetic instabilities \cite{Mat98.1}.
At the same time, the superconducting phases found
in rare earth intermetallic compounds have rich and diverse properties.
It is therefore natural to suspect that the global phase diagram for the heavy fermions
with its various magnetic transitions will have implications for the emergence of superconductivity.
That antiferromagnetic spin fluctuations promote unconventional superconductivity
has been suggested ~\cite{Scalapino86,Varma86} 
soon after the discovery of unconventional superconductivity
in CeCu$_2$Si$_2$ by F.~Steglich~\cite{Ste79.1}.
 Inelastic neutron scattering spectroscopy of the normal and superconducting state
of CeCu$_2$Si$_2$ near quantum criticality showed that the gain in exchange energy across the transition
is an order of magnitude larger than the condensation energy \cite{Stockert-natphys11}. 
(Related conclusion has also been reached in CeCoIn$_5$ \cite{Stock.08}.)
While establishing that the 
magnetism drives the formation of superconductivity, it also implies that 
a  correspondingly large kinetic energy is lost across the superconducting transition.
This has been interpreted  \cite{Stockert-natphys11,Stockert-jpsj12} 
in terms of a Kondo-destruction energy scale $E^*_{\mbox{\tiny loc}}$ being 
nonzero but quite small (Fig.~\ref{fig:SC-CeCu2Si2}): this allows the further reduction of the Kondo-singlet amplitude by superconductivity
to transfer substantial amount of Kondo-resonance spectral weight to higher energies, causing a large 
loss of the Kondo screening energy which is counted as a part of the kinetic energy.
Indeed, there has been considerable evidence by now that the low-energy and low-temperature properties 
are consistent with the SDW type of QCP \cite{Gegenwart.98,Arndt_prl.11}, but the dynamics above
a relatively low temperature ($\sim 1$ K) appears to be consistent with the non-SDW, $\omega/T$,
scaling form \cite{Arndt_prl.11}; this suggests that $E^*_{\mbox{\tiny loc}}$ at the QCP is on the order of $1$ K.
In recent years, STM studies have been carried out in a number of heavy-fermion systems
\cite{Sch10.2, Ernst-11, Aynajian-nature12}.
We anticipate that the kinetic energy loss can be estimated through STM measurements of the single-particle 
spectral function both in the normal and superconducting states.

\begin{figure}[t!]
   \centering
   \includegraphics[width=5.0in]{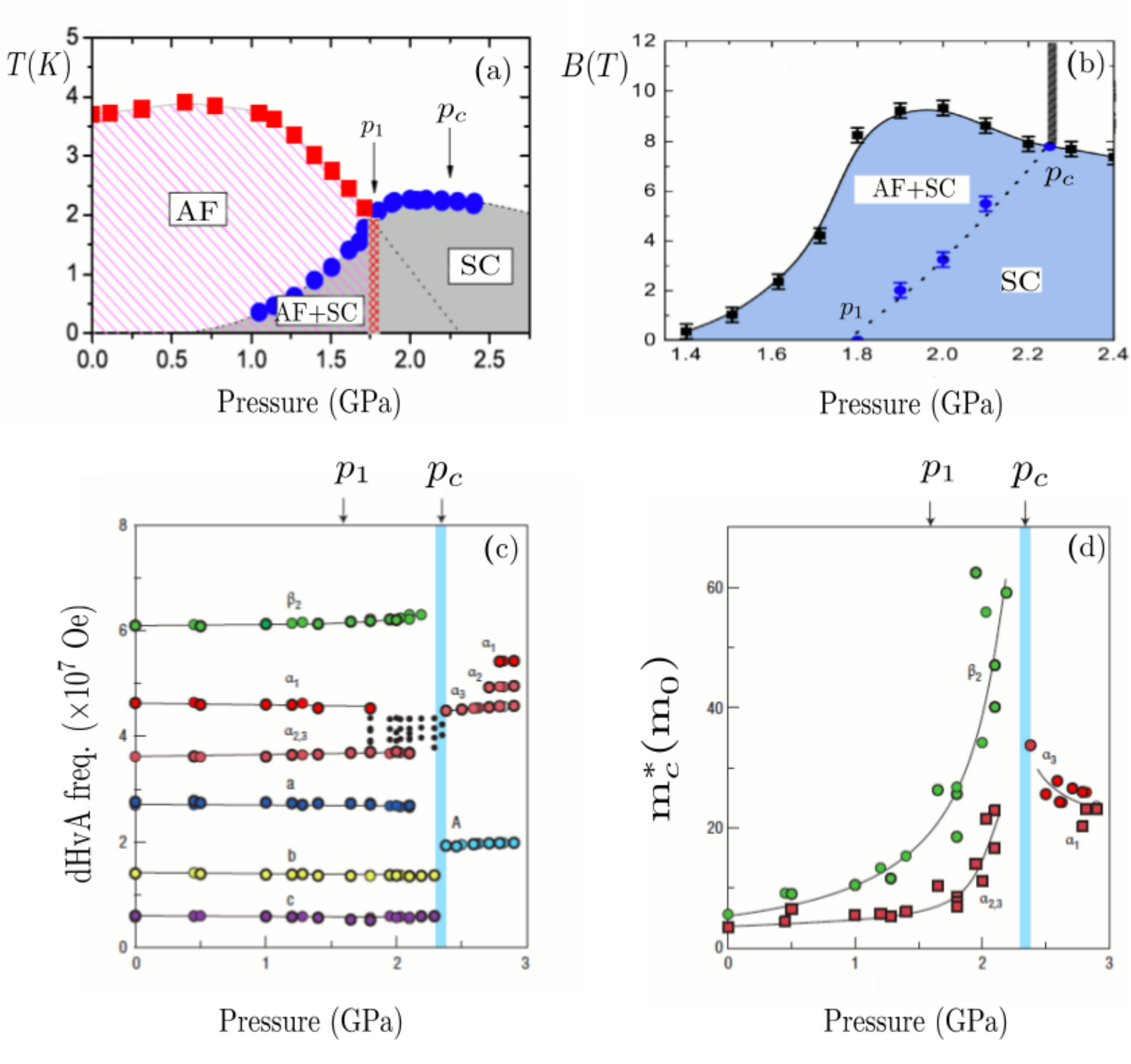}
   \caption{(Color online) Phase diagram of CeRhIn$_5$ in the $T$-$p$ plane at zero field (a)
   and in the $B$-$p$ plane close to zero temperature (b); figures adapted from Ref.~\cite{park-nature06}.
   Also shown is the dHvA measurements as a function of pressure, showing a sudden jump of dHvA frequencies 
   (c), which indicates a corresponding jump of the Fermi surface, and a tendency of divergence in the effective 
   mass (d); figures adapted from Ref.~\cite{Shi05.1}.
      }
   \label{fig:SC-CeRhIn5}
\end{figure}

The evidence for the Kondo destruction quantum criticality determining superconductivity is the most direct in
CeRhIn$_5$ (Refs.~\cite{park-nature06,Shi05.1}). In this compound under pressure, AF order is weakened 
and eventually gives way to superconductivity \cite{Hegger.00},
as shown in Fig.~\ref{fig:SC-CeRhIn5}(a);  $T_c \approx 2.3$ K is high in that it is about 10\% of the bare Kondo
temperature. Applying a magnetic field suppresses superconductivity and uncovers an AF QCP, as seen in 
Fig.~\ref{fig:SC-CeRhIn5}(b). Across this QCP, the Fermi surface experiences a sudden jump 
[Fig.~\ref{fig:SC-CeRhIn5}(c)]. This provides evidence for Kondo destruction, which is corroborated 
by the divergence 
of the effective mass [Fig.~\ref{fig:SC-CeRhIn5}(d)]. All these suggest that the high-temperature superconductivity
in CeRhIn$_5$ originates from local quantum criticality \cite{park-nature06}.

%%% Figure 10%%%%
%\begin{figure}[htbp]
%%%%%%%%%

\section{Spin-orbit coupling and topological phases}

\subsection{Global phase diagram of Kondo insulators}

A generalization to the commensurate conduction-electron filling of 
$x=1$ leads to the corresponding global phase diagram for Kondo insulators
\cite{Yam10.1},
 shown
in Fig.~\ref{fig:gpd_ki}.
As discussed elsewhere~\cite{Si.13.1}, various material families
could be considered as candidates for inducing transitions between these different phases.

\subsection{Topological phases and their transitions to magnetic and Kondo states}

\begin{figure}[t!]
   \centering
%\vskip 0.5 cm
\includegraphics[width=4.2in]{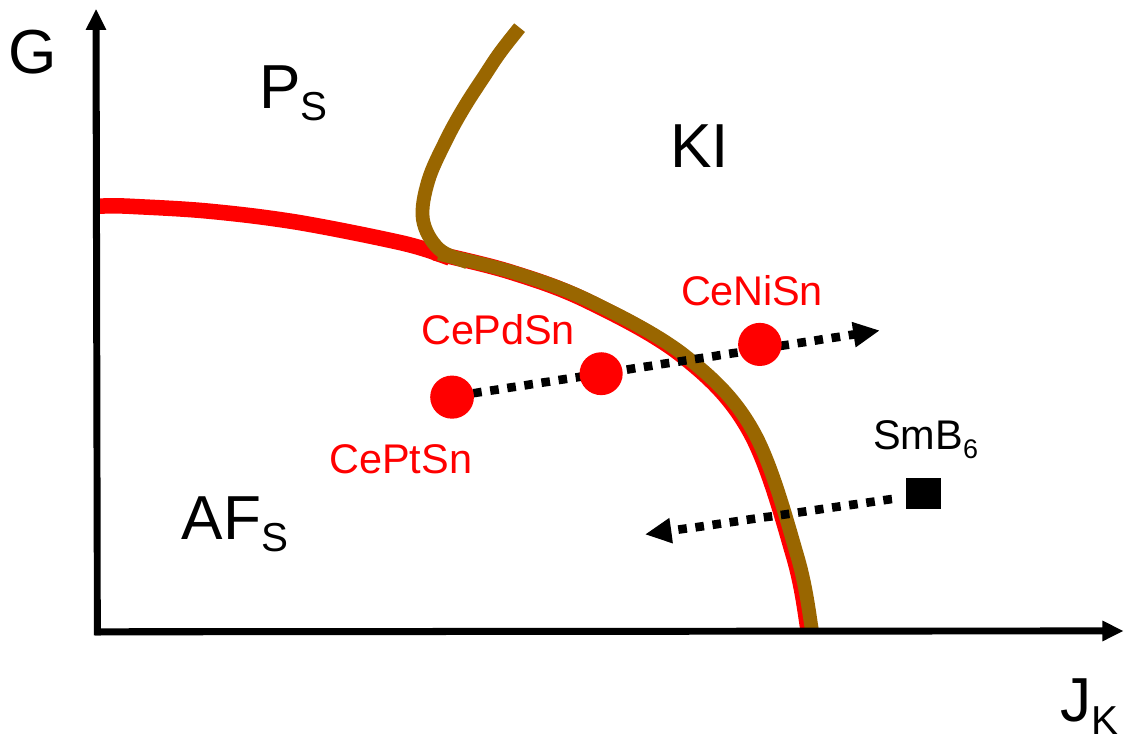}
   \caption{
   (Color online) Global phase diagram of Kondo insulators \cite{Yam10.1}, and representative materials 
   that may be tuned through various transitions \cite{Si.13.1}. Figure adapted from Ref.~\cite{Si.13.1}.
    }
   \label{fig:gpd_ki}
\end{figure}

SmB$_6$ has been the focus of many renewed experiments 
\cite{Wolgast12,Botimer12}. These have followed the suggestion that 
the strong spin-orbit coupling of the 4f-electrons induce non-trivial topology in the heavy-fermion bandstructure,
which turn the Kondo insulator into a topological insulator (TI)~\cite{Dze10.1}. At the present time, there is considerable 
evidence for surface states in SmB$_6$, and whether these are the boundary states of the bulk TI phase remains 
to be established. Still, it is instructive to consider SmB$_6$ as a case in which the bulk
KI gap can be closed by the application of pressure. When that happens, the system becomes metallic and magnetically
ordered~\cite{Bar05.1}, making the trajectory of phase transition to be likely along the dashed line shown in 
Fig.~\ref{fig:gpd_ki}.

%%% Figure 11%%%%
%\begin{figure}[htbp]
%%%%%%%%%%%%

With this in mind, it is intriguing to note the transport evidence for non-Fermi liquid behavior in SmB$_6$ 
under a pressure of about $4$ GPa, 
in the transition regime \cite{Gab03.1}. This suggests that the zero-temperature transition from the KI phase
to the $AF_S$
is (close to being) second order, and the associated QCP underlies the non-Fermi liquid behavior. 
Given the stoichiometric nature of the system, it would then be natural to 
suggest that superconductivity will appear in a similar pressure range as a consequence of quantum
criticality.

Another transition at a Kondo-insulator filling is between a $P_S$ phase and a KI phase \cite{Feng.13}.
This has been studied in
a Kondo lattice model supplemented by a spin-orbit coupling (SOC) for the conduction electrons:
\begin{eqnarray}
H = H_{KL} + H_{soc} (c)
\label{kondo_lattice_soc}
\end{eqnarray}
The details of the Hamiltonian are given in Ref.~\cite{Feng.13}.
The spin-orbit coupling term in this Hamiltonian induces a topological insulator state for the conduction electrons. Because 
of the TI gap of the conduction electrons, the Kondo coupling $J_K$ must be larger than a non-zero threshold value 
$J_K^c$ in order to reach a Kondo insulator phase. A large-N analysis yields a continuous transition between the TI
and Kondo insulator phases. It is likely that magnetic order will also interplay with these phases, 
and studying this effect in the 
model should be very instructive.

%%% Figure 11%%%%
%%%%%%%%%%%%

\subsection{Heavy fermion interfaces}

Such consideration of the spin-orbit coupling also suggests the intriguing possibility of new properties 
at the interface of heavy-fermion heterostructures.
Because of the broken inversion symmetry at the interface, the heavy electrons
in the interface layer should contain an extra SOC of the Rashba type:
${\cal H}_{\rm soc} =                                                              
\sum_{\bf k}  V_{\rm soc} ({\bf n} \times {\bf k}) \cdot {\bf s}({\bf k})$,
where ${\bf k}$ is the wavevector, $V_{\rm soc} $ the                                                                       
spin-orbit
coupling,
${\bf n}$ the unit vector perpendicular to the interface,
and ${\bf s}({\bf k})$
the spin of the electrons with wavevector ${\bf k}$.
For the oxide heterostructures, a Rashba-type SOC with $V_{\rm soc}$ on the order of 
5 meV has been demonstrated\cite{Caviglia.10,Shalom.10}. Such a SOC energy scale will
be competitive against the heavy-fermion energy scales, raising the possibility for topologically 
non-trivial 
superconducting or insulating states at such heavy-fermion interfaces.
Heavy-fermion heterostructures appear to be quite realistic to study. For instance,
 heavy-fermion
superlattices have recently been fabricated
 and studied~\cite{Shi10.1,Matsuda12}.

\section{Summary and Outlook}

With the ever expanding family of heavy fermion materials suitable for studying quantum criticality, there is no 
doubt that new insights will continue to be gained from these systems on general issues of non-Fermi liquid 
behavior and unconventional superconductivity. Here, we have focused our attention on an important theme, 
namely how Kondo destruction influences quantum criticality and the formation of novel phases. 

We have emphasized that quantum criticality that results from Kondo destruction in Kondo lattice system goes beyond the standard 
spin-density-wave type. More generally, we have considered how the physics of Kondo destruction leads to a global 
magnetic phase diagram.  The latter has motivated recent theoretical studies on the interplay between magnetic 
frustration and Kondo screening, as well as the experimental exploration of heavy-fermion compounds with varied dimensionality or geometrical frustration.

We have also provided evidence that superconductivity in some of the canonical heavy-fermion systems is influenced,
or even dominated, by the Kondo-destruction physics of the normal state. Developing a framework to study 
superconductivity in such an unconventional quantum critical setting is a pressing theoretical issue.

Finally, we have discussed how spin-orbit coupling introduces new type of transitions between topological states 
and magnetic or Kondo coherent phases. Pressurizing SmB$_6$ appears to induce such a transition; the tantalizing 
evidence for the (nearly) second-order nature of this transition leads us to suggest that SmB$_6$ under a pressure 
of about 4 GPa might superconduct as a result of the collapsing of the Kondo-insulator gap and the simultaneous 
development of antiferromagnetic order. For related reasons, we have argued that the interface of heavy-fermion heterostructures could be a fertile ground to study the interplay between topological electronic structure, 
magnetism and superconductivity.

%\begin{acknowledgment}

\acknowledgements

We would like to thank J. Dai, L. Deng, X.-Y. Feng, K. Ingersent, J. Wu, J.-X. Zhu and L. Zhu for 
collaborations on the various theoretical aspects covered here,  E. Abrahams, M. Brando,
S. Friedemann, C. Geibel, P. Gegenwart, C. Krellner, S. Paschen, H. Pfau, 
F. Steglich, and S. Wirth for collaborations on the quantum criticality in YbRh$_2$Si$_2$, and S. Paschen for collaborations
on the phase diagram of Ce$_3$Pd$_{20}$Si$_6$ as well as instructive discussions 
on Kondo insulators. This work has been supported in part by
the U.S. Army Research
Office under Grant No. W911NF-13-1-0202 (E.M.N.),
the NSF Grant No. DMR-1309531 and the Robert A. Welch Foundation Grant No.  C-1411.
P.G. acknowledges the support of the NSF Cooperative Agreement No.DMR- 0654118, the State of Florida
and the U. S. Department of Energy.

%\end{acknowledgment}

%\bibliographystyle{jpsj}
%\bibliography{sces13}

\end{document}